\documentclass[12pt,a4paper]{article}
\usepackage{epsf,epsfig,epstopdf,rotating}
\begin{document}
\begin{flushright}
{SINP MSU 2014-2/885\\ 
\hskip 0.5cm\\
January 2014}
\end{flushright}
\vspace{0.5cm}

\begin{center}
{\Large \bf Expectations for the Higgs boson identification at the ILC
\hfill\\}
\end{center}
\vspace{0.5cm}

\begin{center}
{E.~Boos$^{1}$, V.~Bunichev$^{1}$, M.~Dubinin$^{1,2}$, Y.~Kurihara$^2$\\
\hfill\\
{\small \it $^1$Skobeltsyn Institute of Nuclear Physics, Moscow State University}\\
{\small \it 119991, Moscow, Russia} \\
 {\small \it $^2$High Energy Accelerators Research Organization (KEK), Tsukuba, 305-0801 Ibaraki, Japan} }
\end{center}

\vspace{1.0cm}
\begin{center}
{\bf Abstract}
\end{center}
\begin{quote}
{\small Deviations from the standard Higgs sector generated by some new physics at an energy scale $\Lambda$ could be described by an effective $SU(3)_c \times SU(2)_L \times U(1)$ invariant non-renormalizable Lagrangian terms of dimension six. A systematic study of various Higgs boson production channels ($\gamma \gamma$, $ZZ$, $WW$, $b \bar b$, $\tau \bar \tau$) at the International Linear Collider (ILC) in the SM extension by effective operators is carried out. Statistical methods are used to establish a degree of consistency for the standard Higgs sector with the forthcoming data, using the expected ILC accuracies of the Higgs boson production channels. Global fits in the two-parametric anomalous coupling space indicating to possible deviations from the standard Higgs-fermion and Higgs-gauge boson couplings are performed.}
\end{quote}   

\newpage

High precision studies of the electroweak interactions at LEP, SLC and Tevatron \cite{adlos} with the following discovery of a Higgs-like signal at the LHC \cite{signal} and the evidence of a signal at the
Tevatron \cite{signal_tevatron}, confirmed the Standard Model (SM) scheme in the form of $SU(2)_L \times U(1)_Y$ gauge theory with symmetry breaking generated by the Lagrangian which includes one $SU(2)$ doublet of scalar fields. No belief that this simplest possibility is correct naturally results in a desire to analyse theoretically acceptable extensions of the SM Higgs sector and compare their phenomenological consequences with mass, couplings and quantum numbers of the Higgs boson measured precisely in the major production channels at the next-generation $e^+ e^-$ collider.

"Minimal" extensions of the SM by the dimension-five \cite{dim_five} and the dimension-six \cite{buchmuller,dim_six} effective operators, which lead to the rescaling of Higgs couplings, as well as more complicated modifications of Higgs couplings in the MSSM and nonminimal supersymmetric models in the decoupling limit \cite{modifications} commonly demonstrate insignificant (of the order of a few percent) departures of the nonstandard couplings from their SM limit in the most acceptable phenomenolоgical scenarios. Precise measurements are required to study the couplings. Although ATLAS and CMS experiments are sensitive to almost all couplings (except invisible modes and $H\to c \bar c$) which have been analysed recently using the LHC data \cite{lhc_analysis}, the best sensitivity of individual production channels $g(H\to P \bar P)/g(H\to P \bar P)_{SM} -1$ ($P=\gamma,W,Z,\tau,b$) estimated in a model independent way \cite{lhc_estimates} using the {\it production $\times$ decay} approximation is from 5\% to 20\% (these numbers are 1$\sigma$ intervals at the energy 14 TeV with the accumulated luminosity 300 fb$^{-1}$, see \cite{ilc_tdr}). Such accuracy will be undoubtedly sufficient to confirm an agreement of basic properties of the Higgs-like particle (mass, spin and parity) with the properties of the SM Higgs boson, but will be not sufficient to test the deviations at the percent level, inherent for the MSSM in the decoupling limit or models with composite Higgs boson, where the compositeness scale is of the order of 1 TeV.

Clean environment and better signal-to-background ratios of the ILC allows to measure more precisely the couplings of the Higgs boson, which can be observed in all modes including the two-jet decay modes and the invisible modes. Only two production channels are relevant, Higgsstrahlung $e^+ e^- \to ZH$ and vector boson fusion (VBF) $e^+ e^- \to \nu_e {\bar \nu}_e H$. Although it will be not possible to measure directly the total width of the scalar which is about 4 MeV, it can be indirectly determined separating the events with the Higgs recoiling from the Z in the Higgsstrahlung process and measuring directly the branching fractions. Measurement accuracies expected from the ILC experiments (see Table 2.4 in the ILC TDR \cite{ilc_tdr}), estimated from the full detector simulation studies for a realistic accumulated luminosities at $\sqrt{s}=$250 GeV, 500 GeV and 1 TeV are listed in Table 1. Note that the ILC TDR expected accuracies are insignificantly different from the accuracies quoted in the Snowmass Report \cite{snowmass}; an independent simulation of some channels, including invisible modes, can be found in \cite{han}.       

\begin{table}[t]
\begin{center}
\begin{tabular}{|c|c|c|c|c|c|}
\hline
\multicolumn{1}{|c|}{} & \multicolumn {5}{|c|}{$\Delta(\sigma \cdot {\it Br})/(\sigma \cdot {\it Br})$} \\ \hline
$\sqrt{s}$ and ${\cal L}$ & \multicolumn{2}{|c|}{250 fb$^{-1}$ at 250 GeV} & \multicolumn{2}{|c|}{500 fb$^{-1}$ at 500 GeV}& 1 ab$^{-1}$ at 1000 GeV \\ 
($P_{e^-}$,$P_{e^+}$)     & \multicolumn{2}{|c|}{(-0.8,+0.3)}              & \multicolumn{2}{|c|}{(-0.8,+0.3)} & (-0.8,+0.2)           \\ \hline
channel                   &  $ZH$ & $\nu \bar \nu H$ &  $ZH$ & $\nu \bar \nu H$ &  $\nu \bar \nu H$                                 \\ \hline
$H\to b \bar b$ & 1.1\% & 10.5\% & 1.8\% &  0.66\% & 0.47\%     \\
$H\to c \bar c$ & 7.4\% & - & 12.0\% &  6.2\% & 7.6\%     \\
$H\to gg$       & 9.1\% & - & 14.0\% & 4.1\% & 3.1\%       \\
$H\to W W^*$    & 6.4\% & - &  9.2\% & 2.6\% & 3.3\%     \\ 
$H\to \tau^+ \tau^-$ & 4.2\% & - & 5.4\% & 14.0\% & 3.5\%     \\  
$H\to ZZ^*$     & 19.0\% & - & 25.0\% &  8.2\%  & 4.4\%    \\
$H\to \gamma \gamma$ & 29-38\% & - & 29-38\% &  20-26\% & 7-10\%     \\  
$H\to \mu^+ \mu^-$ & 100\% & - & - & - & 32\%    \\ \hline

\end{tabular}
\end{center}
\caption{\label{table_1} Expected accuracies for the Higgs boson production channels at the ILC from \cite{ilc_tdr}.}
\end{table}

\begin{figure}[h]
\begin{minipage}[c]{.50\textwidth}
\includegraphics[width=3in]{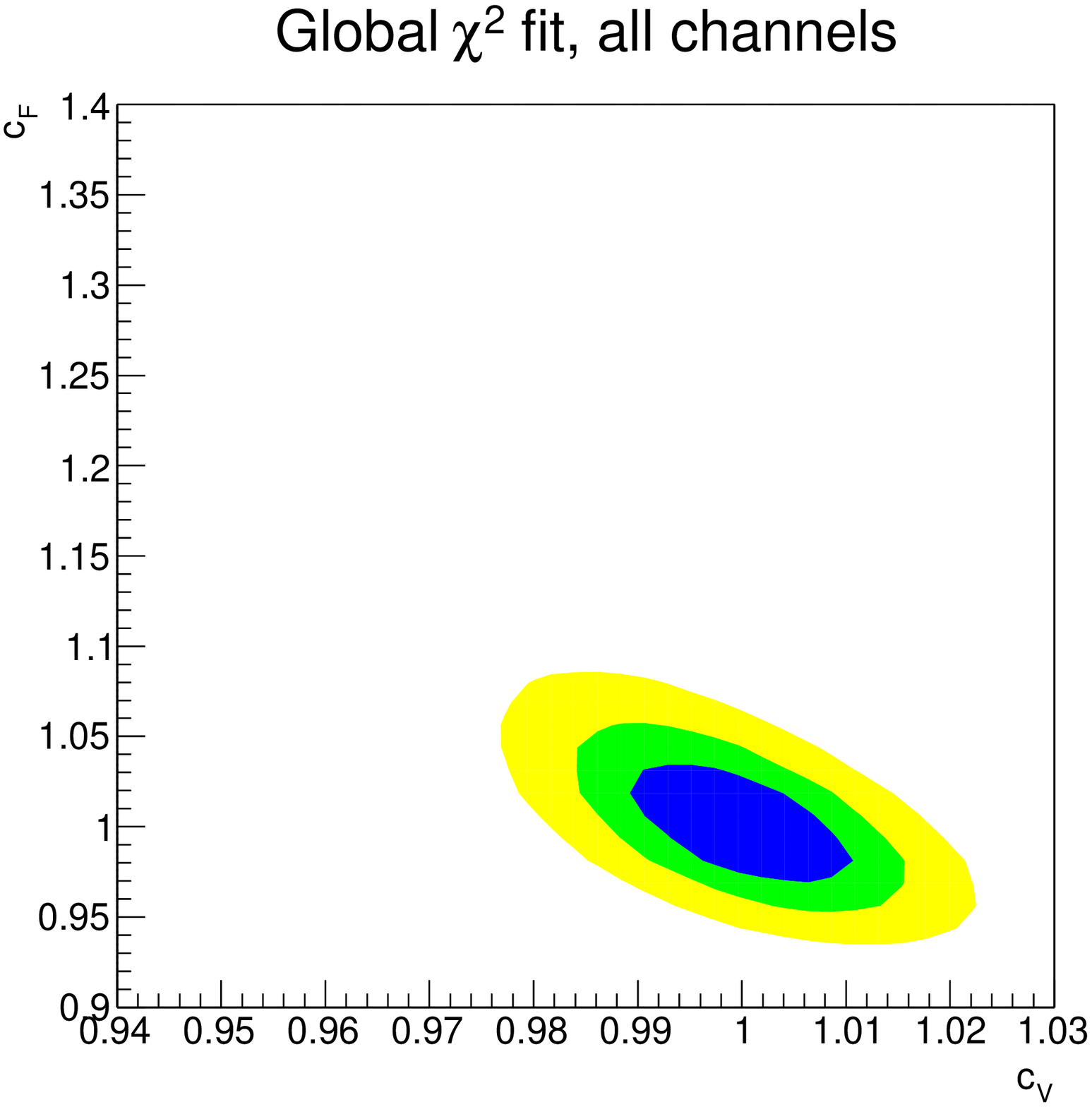}
\begin{center}
{\it (a)}
\end{center}
\end{minipage}
\begin{minipage}[c]{.50\textwidth}
\includegraphics[width=3in]{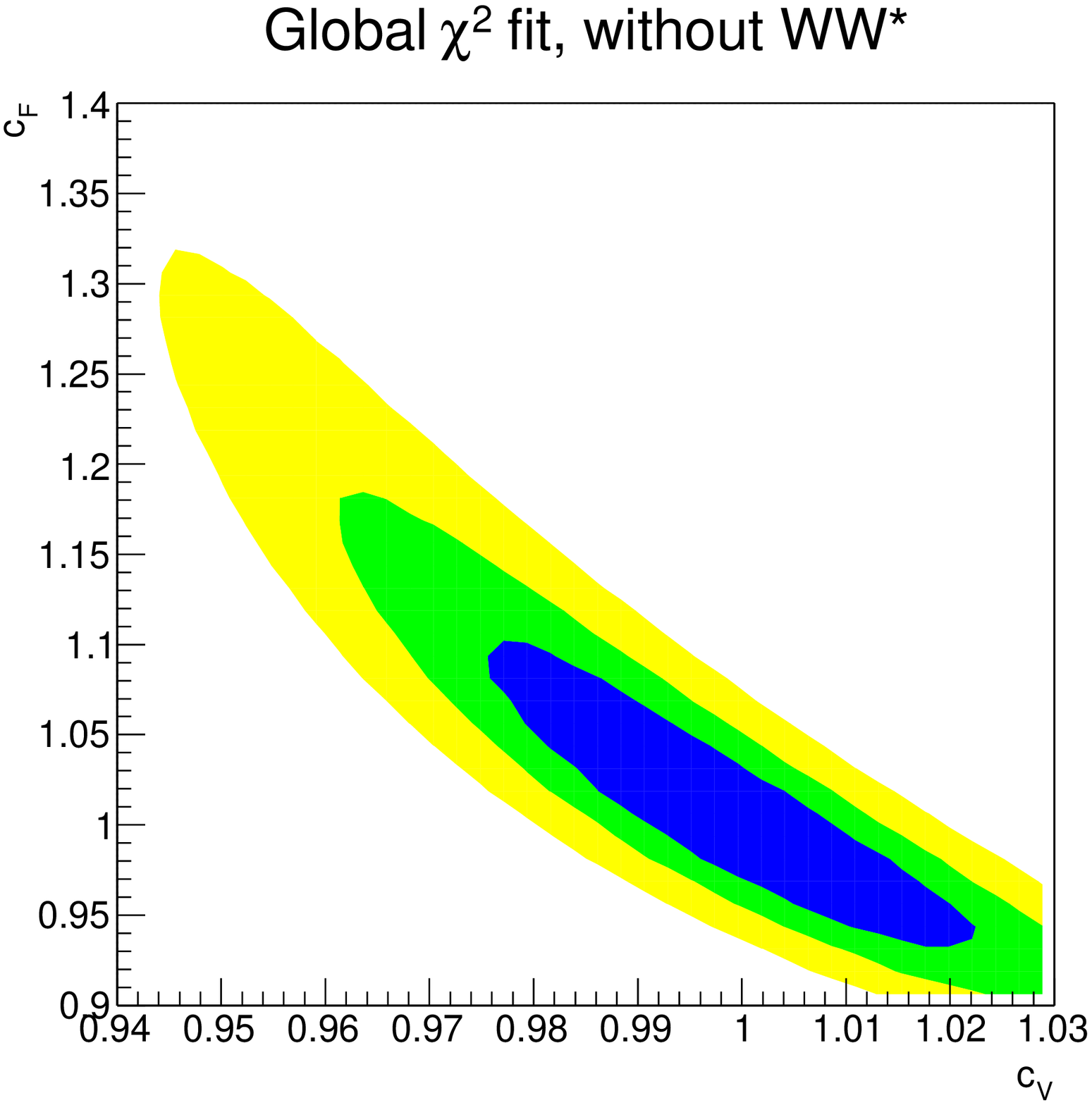}
\begin{center}
{\it (b)}
\end{center}
\end{minipage}
\begin{minipage}[c]{.50\textwidth}
\includegraphics[width=3in]{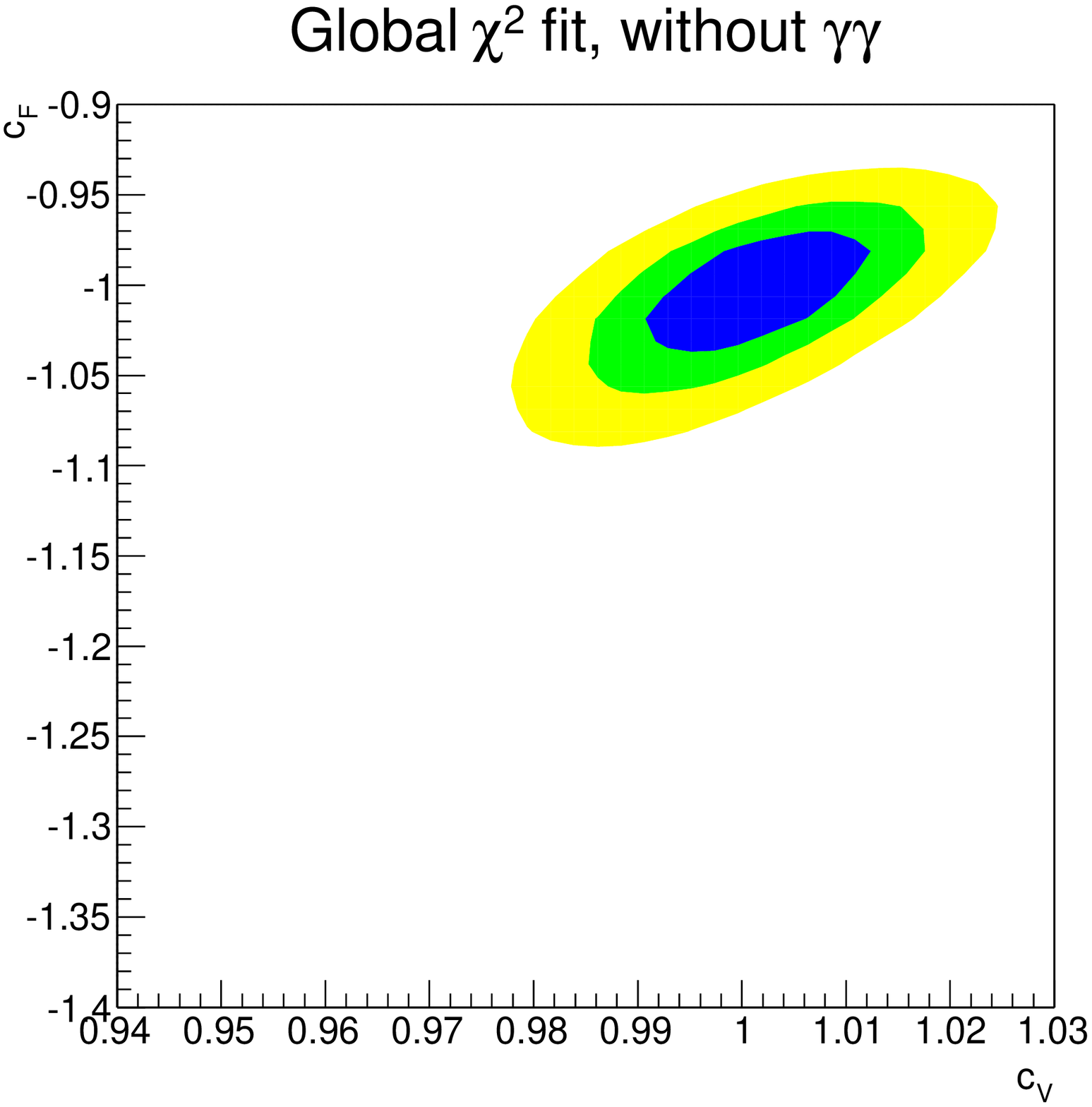}
\begin{center}
{\it (c)}
\end{center}
\end{minipage}
\begin{minipage}[c]{.50\textwidth}
\includegraphics[width=3in]{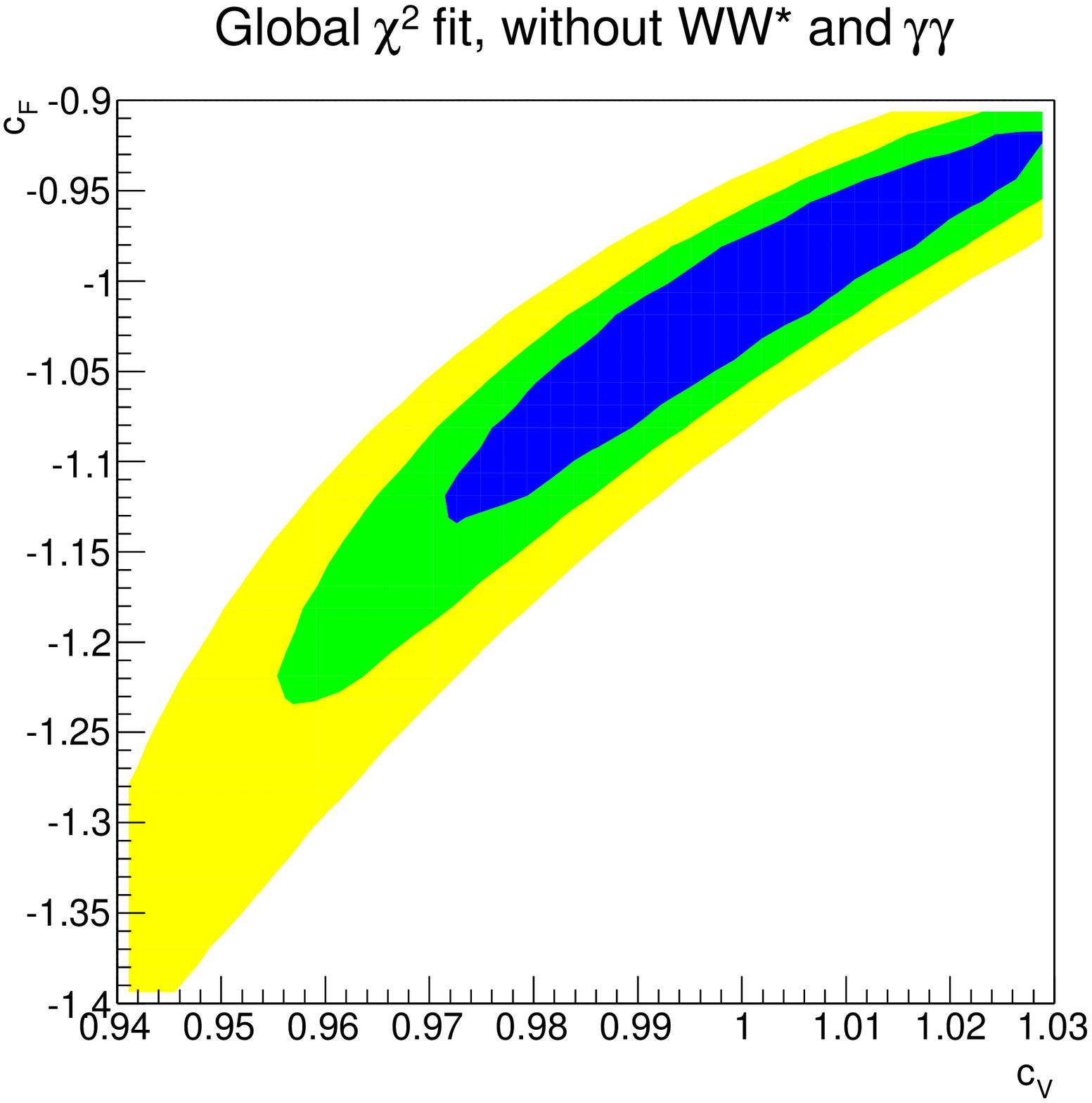}
\begin{center}
{\it (d)}
\end{center}
\end{minipage}
\caption[]{\label{fig_1} Exclusion contours for the combined $\chi^2$ fit in the ($c_V$,$c_F$) plane, $\sqrt{s}=$250 GeV. Blue, green and yellow areas correspond to $\Delta \chi^2=$2.10, 4.61 and 9.21 (CL of the fit is 65\%, 90\% and 99\%), respectively. For the upper row of plots the channel $H \to \gamma \gamma$ is accounted for; its influence on the contours is very small. Lower row of plots shows the acceptable regions which appear without the $H \to \gamma \gamma$ channel. For the plots (b) and (d) the $H\to WW^*$ channel is excluded.}
\end{figure}

\begin{figure}[h]
\begin{minipage}[c]{.50\textwidth}
\includegraphics[width=3.0in]{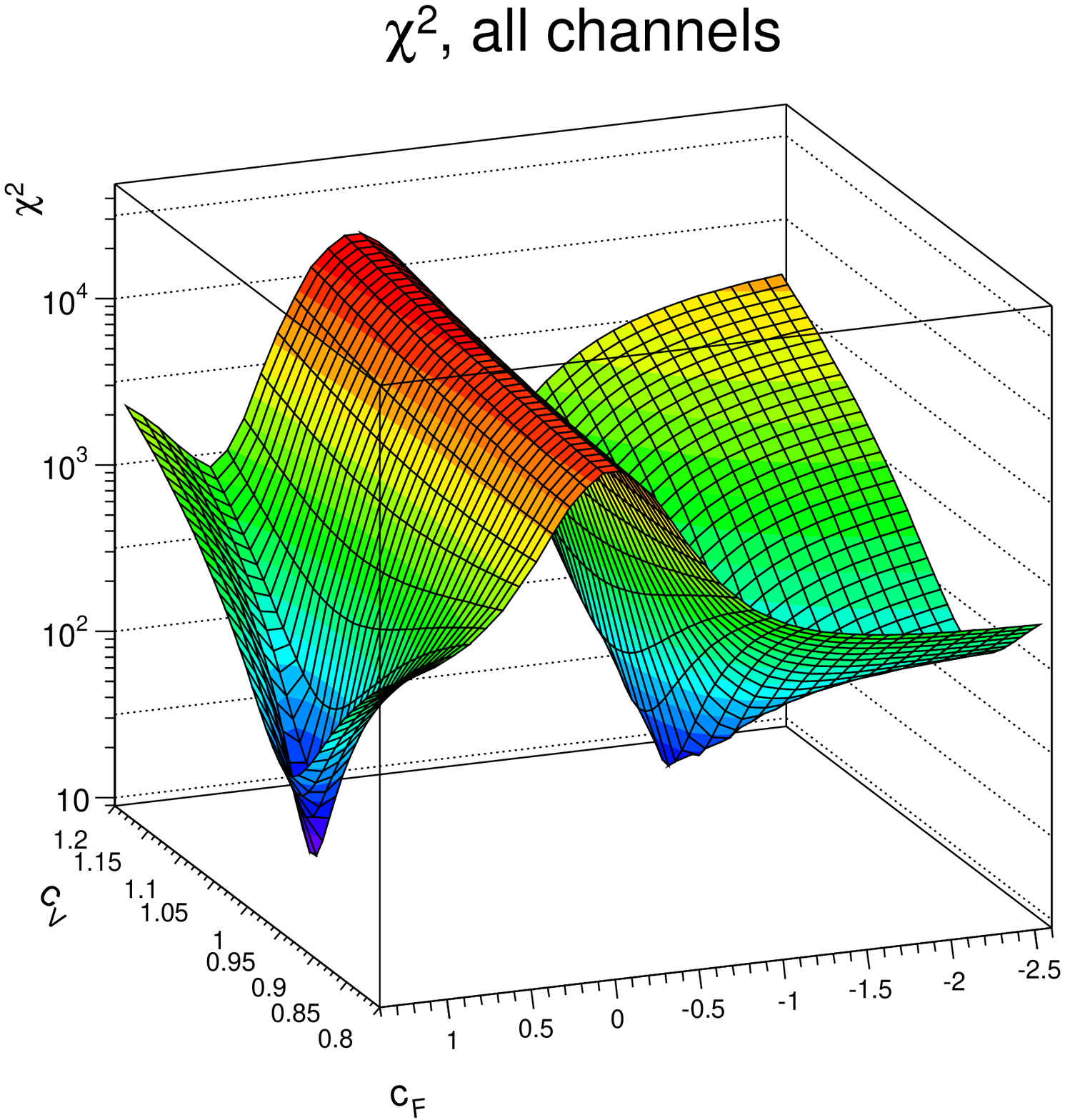}
\begin{center}
{\it (a)}
\end{center}
\end{minipage}
\begin{minipage}[c]{.50\textwidth}
\includegraphics[width=3.0in]{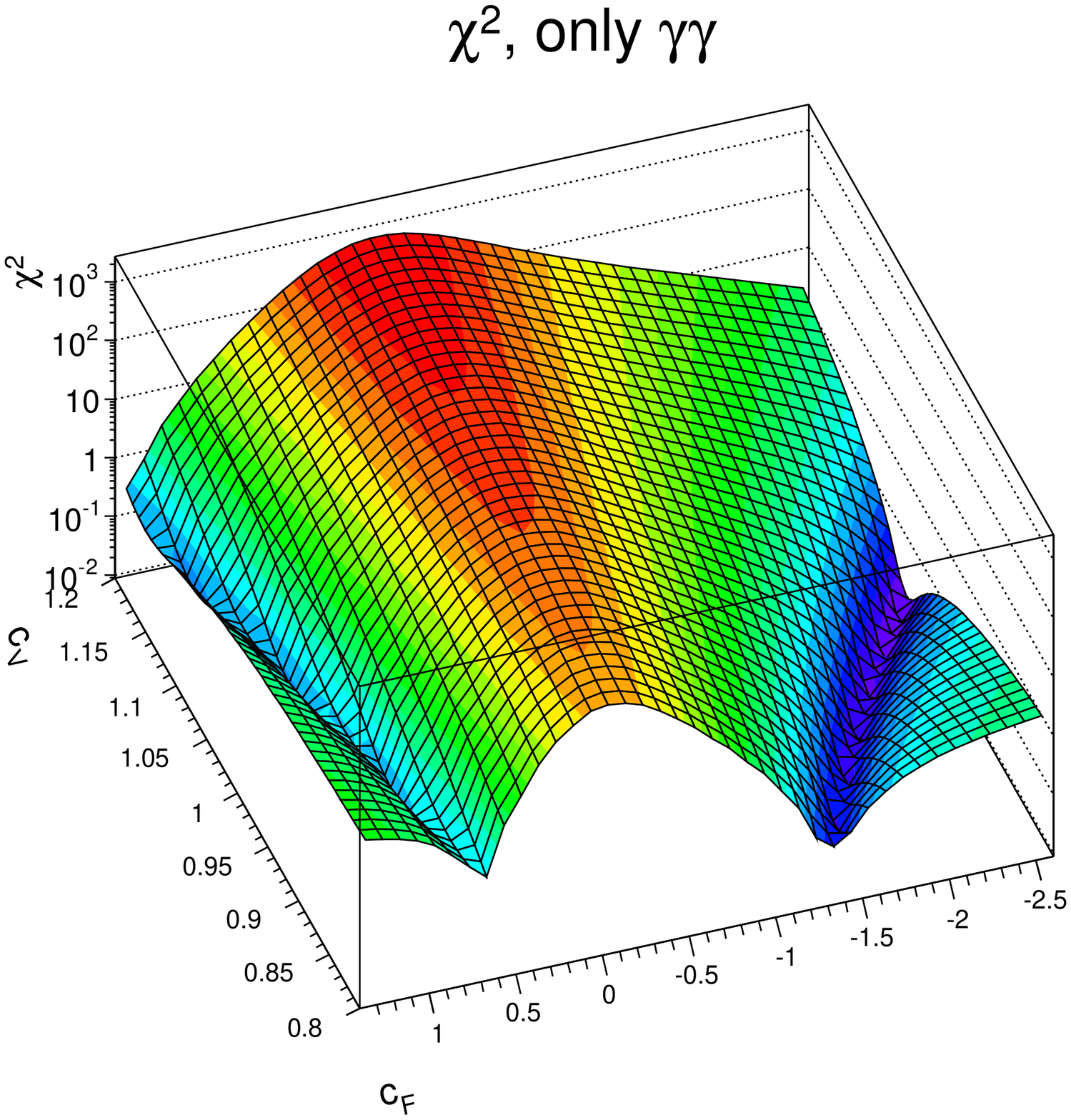}
\begin{center}
{\it (b)}
\end{center}
\end{minipage}

\caption[]{\label{fig_2} Three-dimensional plot of $\chi^2$ as a function of ($c_V$,$c_F$), $\sqrt{s}=$250 GeV, (a) - all channels, (b) - only $H\to \gamma \gamma$ channel. Asymmetric narrow gullies in (b) appear due to the destructive interference of the fermion and the vector boson loop contributions. For this reason the left local minimum of the global $\chi^2$, figure (a), is lower than the right local minimum. The signal strength for all channels is assumed to be equal to 1 and
the individual signal strength errors are very small (see Table 1).}
\end{figure}

\begin{figure}[h]
\begin{minipage}[c]{.50\textwidth}
\includegraphics[width=3in]{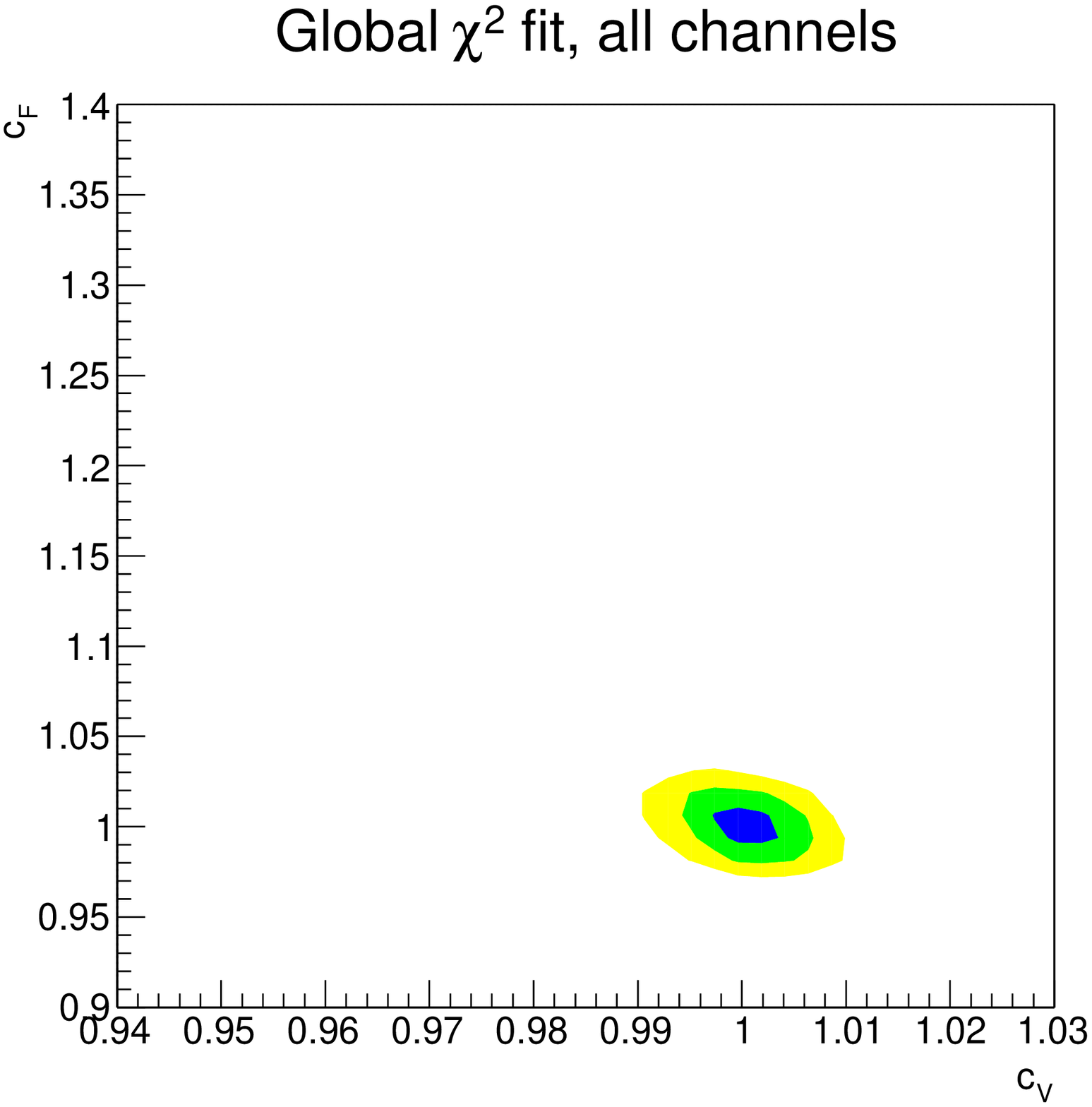}
\begin{center}
{\it (a)}
\end{center}
\end{minipage}
\begin{minipage}[c]{.50\textwidth}
\includegraphics[width=3in]{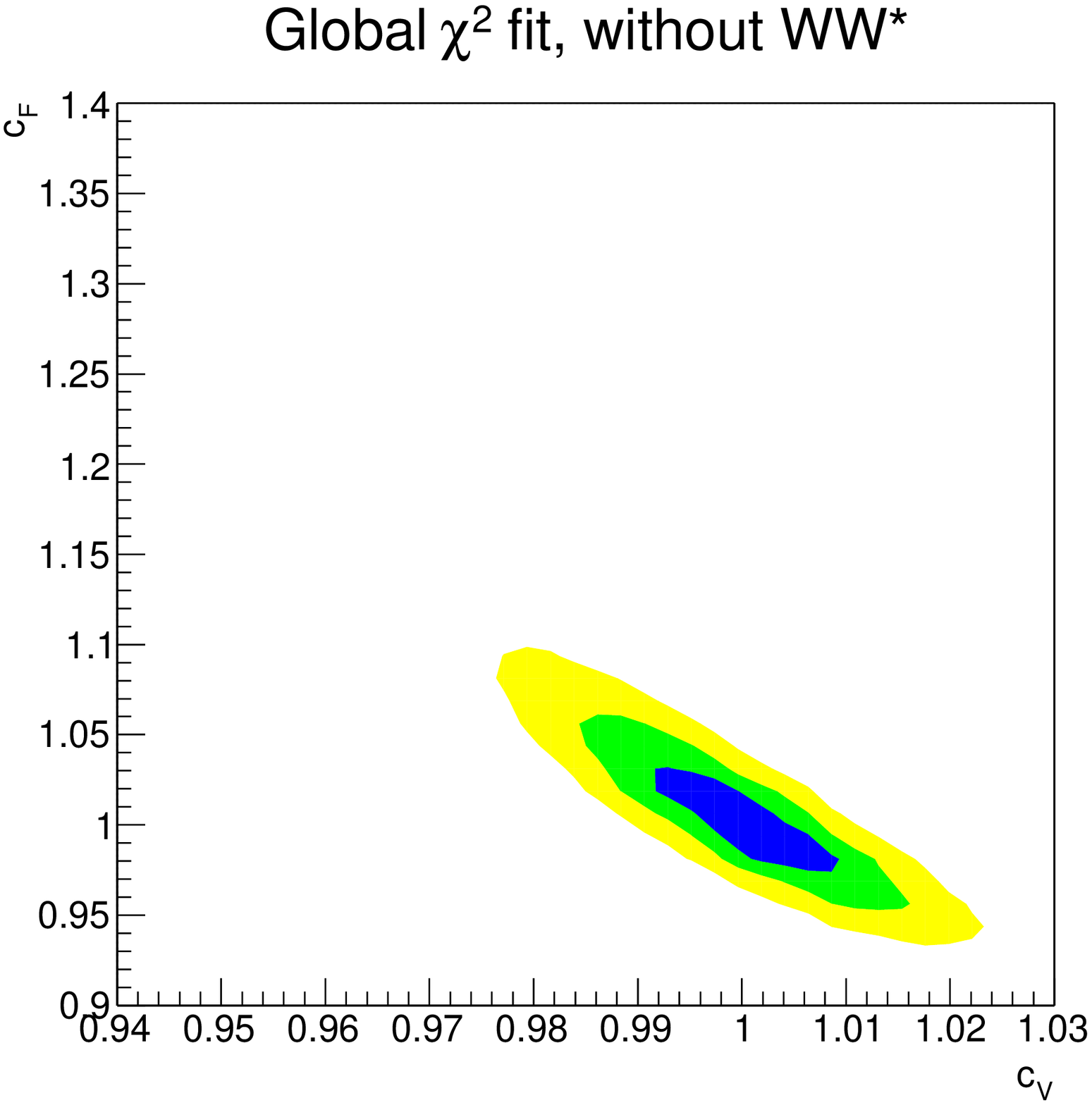}
\begin{center}
{\it (b)}
\end{center}
\end{minipage}
\begin{minipage}[c]{.50\textwidth}
\includegraphics[width=3in]{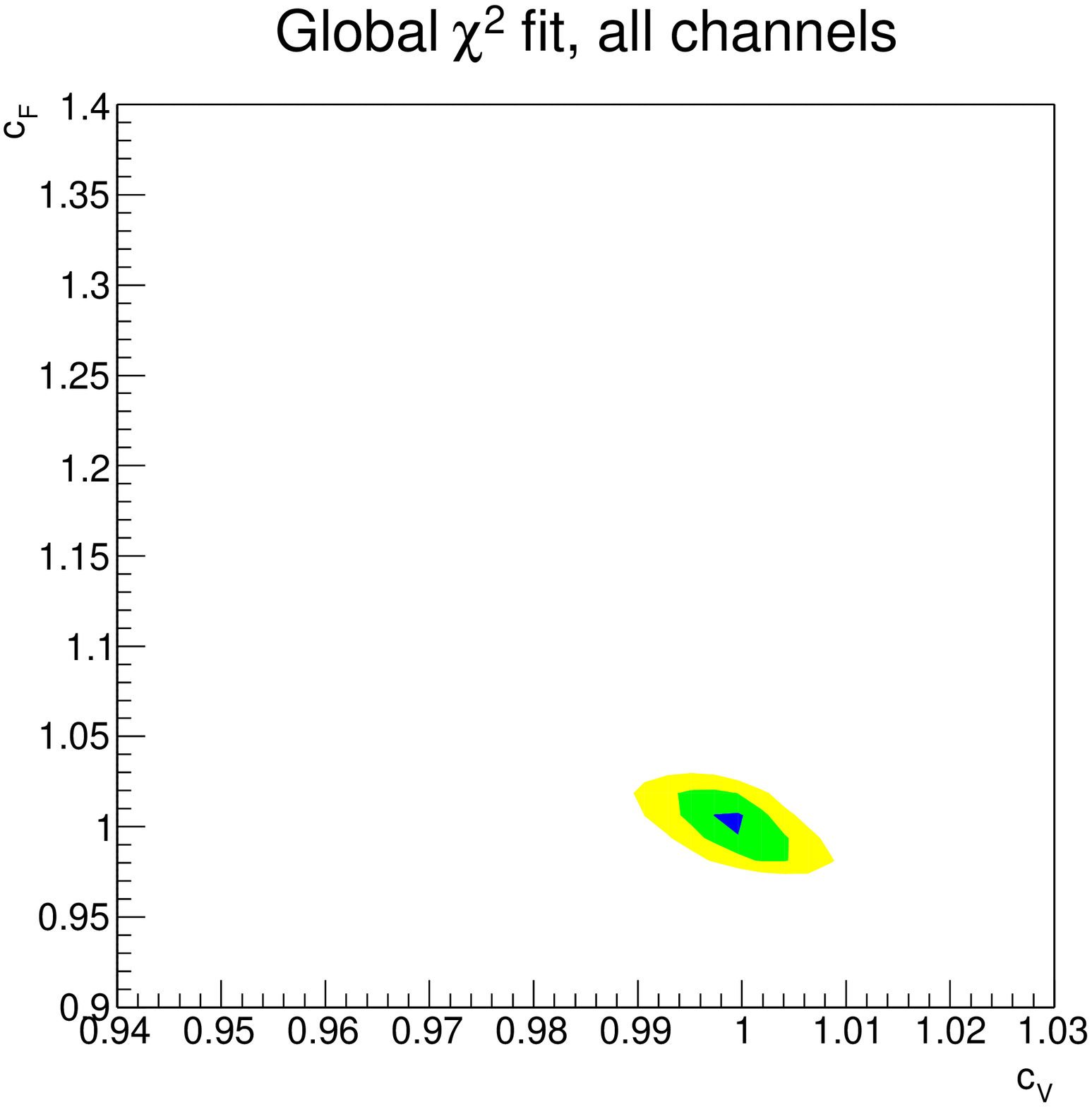}
\begin{center}
{\it (c)}
\end{center}
\end{minipage}
\begin{minipage}[c]{.50\textwidth}
\includegraphics[width=3in]{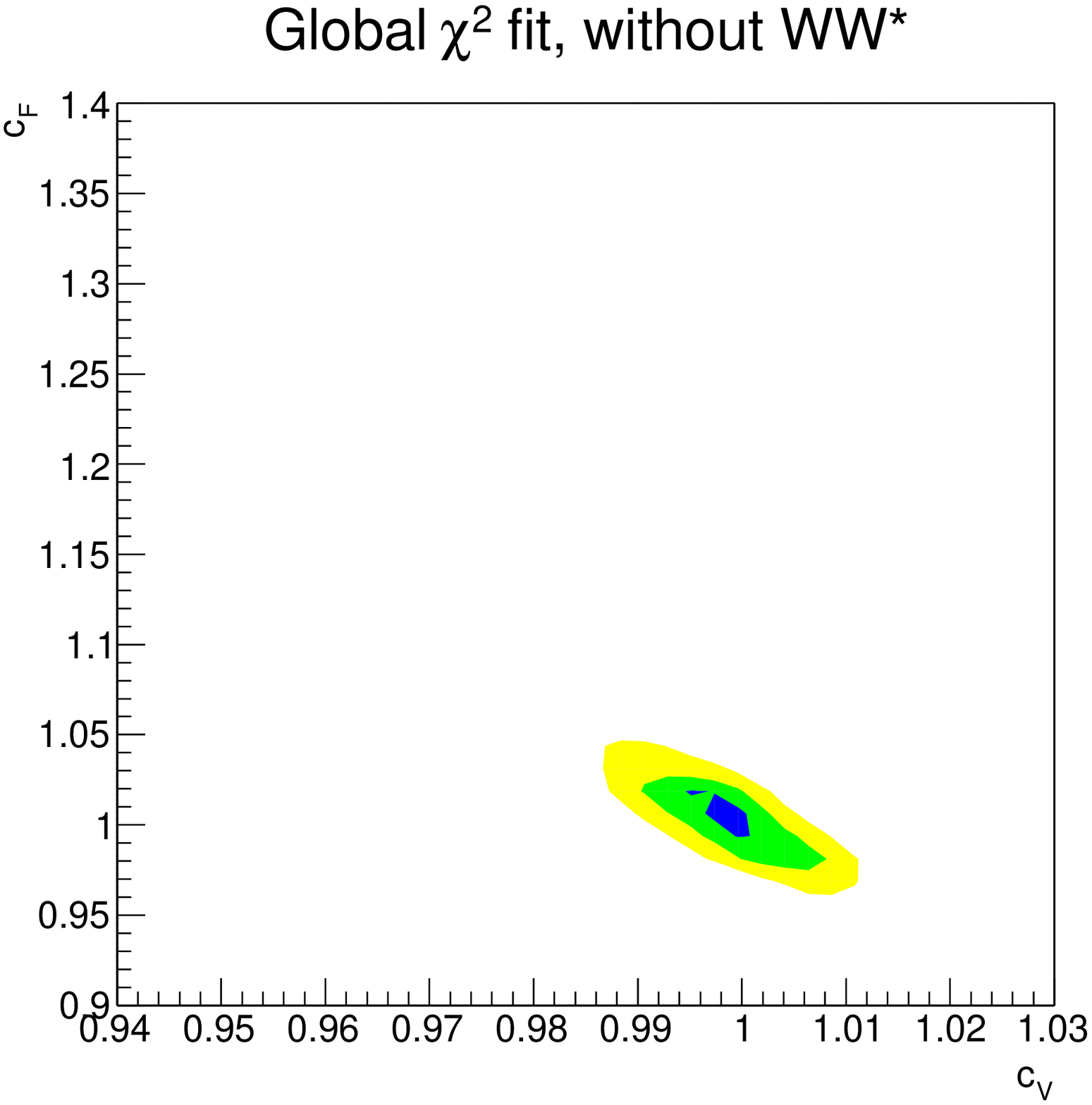}
\begin{center}
{\it (d)}
\end{center}
\end{minipage}
\caption[]{\label{fig_3} Exclusion contours for the combined $\chi^2$ fit in the ($c_V$,$c_F$) plane, $\sqrt{s}=$500 GeV (upper row of plots) and $\sqrt{s}=$1000 GeV (lower row of plots). Blue, green and yellow areas correspond to $\Delta \chi^2=$2.10, 4.61 and 9.21, respectively.  For the plots (b) and (d) the $H\to WW^*$ channel is excluded. }
\end{figure}

In this letter we are using an extension of the SM by the dimension-six effective operators \cite{own} which has been developed for identification of the Higgs-like particle at the LHC. A set of the dimension-six operators in the Buchmueller-Wyler basis \cite{buchmuller} modified by the subtraction of v.e.v.: $\Phi^\dagger \Phi \to \Phi^\dagger \Phi-v^2/2$ (see \cite{passarino}) to avoid undesirable mixing in the gauge field kinetic terms, which is strongly constrained by precision electroweak data, can be reduced to a restricted set of five fermion-Higgs and vector boson-Higgs operators $O_{t \Phi}$, $O_{b \Phi}$, $O_{\tau \Phi}$ and $O^{(1)}_\Phi$, $O_{\Phi G}$, disposing the tensor structure of interaction vertices identical to the SM and dependent only on the two anomalous couplings $c_V$ (rescales vector boson-Higgs vertices) and $c_F$ (rescales fermion-antifermion-Higgs vertices). The anomalous couplings $C_n$, $C_{mn}$ in front of the dimension-six operators $O_n$, $O_{mn}$ are conformably redefined \cite{own}, for example 
\begin{eqnarray*}
c_F= 1 + C_{t\Phi}\cdot\frac{v^2}{\Lambda^2}, \hskip 0.5cm
c_V= 1+\frac{v^2}{2\Lambda^2}\cdot C^{(1)}_{\Phi}, \hskip 0.5cm
c_G= c_F+\frac{6\pi}{\alpha_s}\cdot C_{\Phi G}\cdot \frac{v^2}{\Lambda^2}, \hskip 0.3cm...
\end{eqnarray*}
($v=$246 GeV, $\Lambda$ is the energy scale of new physics). The last equation here means that in the approach under consideration the one-loop $H\to \gamma \gamma$ and $H\to gg$ vertices are "resolved" in the sense that the anomalous parameters $c_V$ and $c_F$ are included in the one-loop effective $c_G$ and $c_\gamma$ which parametrise $H\to gg$ and $H\to \gamma \gamma$ vertices. Such parametrization is different from phenomenological parametrization used by ATLAS and CMS collaborations \cite{hxswg}, where independent anomalous couplings and total widths are simultaneously used. Statistical methods for the signal strength equal to one and the signal strength error defined by the abovementioned ILC TDR signal simulation studies are used to evaluate combined $\chi^2$ fits in the ($c_V$,$c_F$) anomalous coupling plane. Details of the procedure for exclusion contours reconstruction in the ($c_V$,$c_F$) plane are described in \cite{own} (see also \cite{fitting}).  A combination of simultaneous fits for all possible Higgs boson production channels is presented in Fig.\ref{fig_1} and Fig.\ref{fig_3}, where the exclusion contours for $\Delta \chi^2=$2.10, 4.61 and 9.21 are reconstructed at the energies 250, 500 and 1000 GeV. Small signal strength error in the channels $e^+ e^- \to ZH,\nu {\bar \nu} H$ with the following $H\to b {\bar b}$ or $H \to W W^*$ results in the most significant influence of the modes $H\to b {\bar b}$ and $H \to W W^*$ on the combined $\chi^2$ fit. The shape of exclusion contours is most sensitive to the interplay of these two modes, which have opposite behaviour of the cross section dependence $\sigma(c_V,c_F)$ on the anomalous couplings. In order to demonstrate the sensitivity, in Fig.\ref{fig_1}a,b we separately plot the overall fit and the combined fit for all channels excluding $H\to W W^*$ (upper row of plots). Naive expectation that $H\to WW^*$ channel should be more sensitive to $c_V$, than to $c_F$, is delusive since the branching ratios are generally speaking not proportional to partial widths because of an untrival dependence of the total width on the anomalous couplings. Despite a small cross section and low expected precision, the channel $H\to \gamma \gamma$ is very important to exclude the region in the vicinity of ($c_V$,$c_F$)=(1,-1) which is acceptable for LHC global fits (see \cite{own}). In the absence of $H\to \gamma \gamma$ data this region immediately appears for the ILC combined fit, see Fig.\ref{fig_1}, lower row of plots, due to an asymmetric dependence of $\chi^2$($c_V$,$c_F$) on the fermion anomalous coupling $c_F$, illustrated by the three-dimensional 
$\chi^2$ in Fig.\ref{fig_2}a. Asymmetric behaviour with respect to $c_F$ is a consequence of asymmetric behavior of $\chi^2$ for the $H\to \gamma \gamma$ channel taken separately, see Fig.\ref{fig_2}b. At the energies $\sqrt{s}=$500 GeV and $\sqrt{s}=$1000 GeV the role of VBF increases with the cross section growth and smaller signal strength error, as illustrated by Fig.\ref{fig_3}.

Evaluation of the exclusion contours was carried out\footnote{A special regime of 'table calculations' (numerical operations with multidimensional tables) has been implemented in CompHEP version 4.5 \cite{CompHEP}.} for complete gauge invariant sets of diagrams $e^+ e^- \to (HZ$ or $\nu {\bar \nu} H)\to (f {\bar f} f {\bar f}$ or $f \bar f VV$) with the four-fermion and the two-fermion $-$ two vector boson (including $H\to gg$) final states.  Interferences  play a role for LHC fits \cite{own}. For the ILC case, some examples also had been found; e.g. in connection with LEP2 analyses it had been mentioned that the four-fermion state $\nu_e {\bar \nu}_e {b \bar b}$ was formed by both $e^+ e^- \to HZ$ and $e^+ e^- \to \nu_e {\bar \nu}_e H$ interfering production mechanisms, see details at LEP2 and NLC energies in \cite{nnbb}. Untrivial interferences between signal diagrams, not available in the {\it production $\times$ decay} approximation, were found insignificant for our fits, although they could play a role in specific reconstructions of phase space distributions beyond the infinitely small width approximation \cite{lep2}.  

In summary, we analysed possible degree of consistency of the SM Higgs sector with the forthcoming ILC data taking into account the main mechanisms of the Higgs boson production 
$e^+ e^- \to HZ, \nu {\bar \nu} H$ in the framework of the SM extension by dimension-six effective operators.
Combined $\chi^2$ fits at the energies $\sqrt{s}$= 250, 500 and 1000 GeV were carried out and the exclusion regions in ($c_V$,$c_F$) anomalous coupling plane were reconstructed using an expected accuracies for the cross section measurements at the ILC, based on full detector simulation studies. The channel $H\to \gamma \gamma$ is important to exclude the region of anomalous coupling plane close to ($c_V$,$c_F$)=(1,-1).

\begin{center}
{\bf Acknowledgements}
\end{center}

The work of E.B., V.B. and M.D. was partially supported by RFBR grant 12-02-93108 and NSh grant 3042.2014.2.

\newpage

\newpage

\newpage

\end{document}